\documentclass[twocolumn,showpacs,preprintnumbers,amsmath,amssymb]{revtex4}

\usepackage{graphicx}
\usepackage{dcolumn}
\usepackage{bm}
\usepackage{color}
\newcommand{\sqrtsnn}{\sqrt{s_{\mbox{\tiny{\it{NN}}}}}}

\newcommand{\AaAa}{{$A\mbox{+}A$ }}

\newcommand{\PbPb}{{Pb+Pb }}

\newcommand{\pT}{{$p_T$ }}

\newcommand{\hydjet}{{\sc hydjet++ }}
\newcommand{\hj}{{\sc hydjet++}}
\newcommand{\fastmc}{{\sc fastmc }}

\newcommand{\pyquen}{{\sc pyquen }}

\newcommand{\GeVc}{${\rm GeV}/c$}

\begin{document}

\title{ Jets and elliptic flow correlations at low and high 
       transverse momenta in ultrarelativistic \AaAa collisions }

\author{L.V.~Bravina}
\altaffiliation[Also at ]{
Skobeltsyn Institute of Nuclear Physics, Lomonosov Moscow State 
University, RU-119991 Moscow, Russia 
\vspace*{1ex}}
\affiliation{
Department of Physics, University of Oslo, PB 1048 Blindern, 
N-0316 Oslo, Norway
\vspace*{1ex}}
\author{G.Kh.~Eyyubova}
\affiliation{
Skobeltsyn Institute of Nuclear Physics, Lomonosov Moscow State 
University, RU-119991 Moscow, Russia 
\vspace*{1ex}}
\author{V.L.~Korotkikh}
\affiliation{
Skobeltsyn Institute of Nuclear Physics, Lomonosov Moscow State 
University, RU-119991 Moscow, Russia 
\vspace*{1ex}}
\author{I.P.~Lokhtin}
\affiliation{
Skobeltsyn Institute of Nuclear Physics, Lomonosov Moscow State 
University, RU-119991 Moscow, Russia 
\vspace*{1ex}}
\author{S.V.~Petrushanko}
\affiliation{
Skobeltsyn Institute of Nuclear Physics, Lomonosov Moscow State 
University, RU-119991 Moscow, Russia 
\vspace*{1ex}}
\author{A.M.~Snigirev}
\altaffiliation[Also at ]{
Bogoliubov Laboratory of Theoretical Physics, Joint 
Institute for Nuclear Research, RU-141980 Dubna, Russia
\vspace*{1ex}}
\affiliation{
Skobeltsyn Institute of Nuclear Physics, Lomonosov Moscow State 
University, RU-119991 Moscow, Russia 
\vspace*{1ex}}
\author{E.E.~Zabrodin}
\altaffiliation[Also at ]{
Department of Physics, University of Oslo, PB 1048 Blindern, 
N-0316 Oslo, Norway
\vspace*{1ex}}
\affiliation{
Skobeltsyn Institute of Nuclear Physics, Lomonosov Moscow State 
University, RU-119991 Moscow, Russia 
\vspace*{1ex}}

\begin{abstract}
Data from the Large Hadron Collider on elliptic flow correlations 
at low and high $p_T$ from \PbPb collisions at $\sqrtsnn = 5.02$~TeV 
are analyzed and interpreted in the framework of the \hydjet model. 
This model allows us to describe simultaneously the region of both 
low and high transverse momenta and, therefore, to reproduce 
the experimentally observed nontrivial centrality dependence 
of elliptic flow correlations. The origin of the correlations between
low and high-\pT flow components in peripheral lead-lead collisions is
traced to correlations of particles in jets. 

\end{abstract}

\maketitle

\section{Introduction}
\label{intro}

A number of exquisite and intriguing phenomena, which have never been 
systematically studied at the accelerators of the previous generation, 
have been observed since the start of the Relativistic Heavy Ion 
Collider (RHIC) and Large Hadron Collider (LHC) heavy-ion programs. 
Azimuthal anisotropy of multi-particle production is among them as a 
powerful probe of collective properties of a new state of matter, 
quark-gluon plasma (QGP); see, e.g., \cite{Proceedings:2019drx}. It 
is commonly described by the Fourier decomposition of the invariant 
cross section in the form \cite{VoZh_96,PoVo_98} 
\begin{eqnarray}
\displaystyle
\label{eq:1}
E\frac{d^3N}{dp^3} &=&\frac{d^2N}{2\pi p_{T}dp_{T}d\eta} 
\nonumber \\
 &\times & \left\{
1+2\sum\limits_{n = 1}^\infty v_{n}(p_{T},\eta)  
\cos{ \left[ n(\varphi -\Psi_{n}^{PP}) \right] }
\right\},
\end{eqnarray}
where $p_{T}$ is the transverse momentum, $\eta$ is the pseudorapidity, 
$\varphi$ is the azimuthal angle with respect to the participant plane 
$\Psi_{n}^{PP}$ of $n$th order, and $v_{n}$ are the Fourier 
coefficients:
\begin{equation}
\label{eq:2}
v_n =  \langle\langle \cos{ \left[ n(\varphi -\Psi_{n}^{PP}) \right]}
\rangle\rangle.
\end{equation}
The averaging in the last equation is performed over 
all particles in a single event and over all events.

The second harmonic, $v_2$, typically referred to as elliptic flow, 
is the most thoroughly investigated one (for review see 
\cite{HeSn_13} and references therein), because the eccentricity is 
the most pronounced coefficient in the initial conditions. It 
directly relates the anisotropic shape of the overlap region of the 
colliding nuclei to the corresponding anisotropy of the outgoing 
momentum distribution. At relatively low transverse momenta, 
$p_{T} < 3 - 4$~\GeVc, the azimuthal anisotropy results from a 
pressure-driven anisotropic expansion of the created matter, with 
more particles emitted in the direction of the largest pressure 
gradients \cite{Ollitrault:1992bk}. At higher transverse momenta, 
this anisotropy is understood to arise from the 
path-length-dependent energy loss of partonic jets as they traverse 
the matter, with more jet particles emitted in the direction of 
shortest path length \cite{Gyulassy:2000gk}. The correlations 
between soft and hard contributions to anisotropic flow have 
attracted a lot of attention; see, e.g., 
\cite{ASW_05,jia_13,NH_16} and references therein.
Considering initial-state collision geometry asymmetries and 
fluctuations, in Ref.~\cite{NH_16} the authors were able to 
reproduce simultaneously the measurements of nuclear modification 
factor $R_{AA}$ and elliptic flow $v_2$, that was a serious 
problem for most quenching models.

Recently, interesting correlations between $v_2$ values at high 
and low transverse momenta for different centralities were 
reported by the CMS Collaboration \cite{CMS_cor}. These 
correlations can also be seen in the ATLAS data \cite{ATLAS_flow}. 
In the present paper we analyze and interpret this intriguing 
experimental observation within the \hydjet model, which allows 
us to perform such analysis due to its remarkable feature, namely 
the presence of soft and hard physics simultaneously. 

The paper is organized as follows. Basic features of the model 
are sketched in Sec.~\ref{sec2}. Here the origin of elliptic 
flow in the model, and interplay between the soft and hard 
physics, are discussed. Section \ref{sec3} presents the results 
of model calculations of hadronic elliptic flow in \PbPb 
collisions at $\sqrtsnn = 5.02$~TeV. The calculations are in 
fair agreement with the experimental data, and the role of jets 
is clarified. Finally, conclusions are drawn in Sec.~\ref{sec4}.

\section{\hydjet model}
\label{sec2}

The model \hydjet \cite{Lokhtin:2008xi} is a widely used event 
generator, which describes successfully the large number of 
physical observables measured in heavy-ion collisions during 
RHIC and LHC operation. Among them are centrality and 
pseudorapidity dependence of inclusive charged particle 
multiplicity, transverse momentum spectra and $\pi^\pm \pi^\pm$ 
correlation radii in central \PbPb collisions~\cite{Lokhtin:2012re}, 
momentum and centrality dependence of elliptic and higher-order 
harmonic coefficients 
\cite{prc_09,prc_13,Bravina:2013xla,prc_14,prc_17}, flow 
fluctuations \cite{Bravina:2015sda}, angular dihadron 
correlations \cite{Eyyubova:2014dha}, forward-backward 
multiplicity correlations \cite{Zabrodin:2020cql}, jet quenching 
effects \cite{jet1,jet2}, and heavy meson production 
\cite{Lokhtin:2016xnl,Lokhtin:2017rvj,Lokhtin:2019lfm}. 
Details of the model can be found in the \hydjet 
manual~\cite{Lokhtin:2008xi}.

The event generator includes two independent components: the 
soft, hydro-type state and the hard state resulting from 
in-medium multiparton fragmentation. The soft component is the 
thermal hadronic state generated on the chemical and thermal 
freeze-out hypersurfaces prescribed by the parametrization of 
relativistic hydrodynamics with preset freeze-out conditions. 
It represents the adapted version of the event generator 
\fastmc~\cite{Amelin:2006qe,Amelin:2007ic}. Particle 
multiplicities are calculated using the effective thermal 
volume approach and Poisson multiplicity distribution around 
its mean value, which is supposed to be proportional to the 
number of participating nucleons for a given impact parameter 
in an \AaAa collision. 

To simulate the elliptic flow effect, the hydro-inspired 
parametrization for the momentum and spatial anisotropy of soft 
hadron emission source is 
implemented~\cite{Lokhtin:2008xi,Wiedemann:1997cr}.
Note that there are two parameters which govern the strength and the 
direction of the elliptic flow in the original \hydjet
version~\cite{Lokhtin:2008xi}. The first one is the spatial anisotropy
$\epsilon(b)$. It is responsible for the elliptic modulation of the 
final freeze-out hypersurface at a given impact parameter $b$. The
second one is the momentum anisotropy $\delta(b)$, dealing with the 
modulation of the flow velocity profile. One can treat these two 
parameters as independent ones and, therefore, adjust them separately 
for each centrality by comparing to data. Although it provides better 
agreement with the data, this procedure leads to significant increase 
of the parameters to be tuned. Therefore, for the sake of simplicity 
we opted for another scenario, which is implemented in the basic 
version of the model. According to it, both parameters are correlated 
through the dependence of the elliptic flow coefficient $v_2$ on both 
$\epsilon(b)$ and $\delta(b)$, obtained in the hydrodynamic
approach~\cite{Wiedemann:1997cr}:
\begin{equation}
\label{v2-eps-delta1}
v_2(\epsilon, \delta) \propto \frac{2(\delta-\epsilon)}{(1-\delta^2)
(1-\epsilon^{2})}~.
\end{equation}

Because $v_2(b)$ is proportional to the initial ellipticity
$\epsilon_0 (b)=b/2R_A$, where $R_A$ is the radius of colliding 
nucleus, the relation between $\epsilon (b)$ and $\delta(b)$ 
reads~\cite{Lokhtin:2008xi}
\begin{equation}
\label{v2-eps-delta2}
\delta = \frac{\sqrt{1+4B(\epsilon+B)}-1}{2B}~, \quad 
B=C(1-\epsilon^2)\epsilon~, \quad \epsilon=k \epsilon_0~.
\end{equation}
Two new parameters $C$ and $k$, entering the last expression, are 
independent of centrality and transverse momentum and, therefore, 
should be obtained from the fit to the experimental data.

In the hard sector the model propagates the hard partons through the 
expanding quark-gluon plasma and takes into account both collisional 
loss and gluon radiation due to parton rescattering. It is based on 
the \pyquen partonic energy loss model~\cite{Lokhtin:2005px}.
The number of jets is generated according to a binomial distribution. 
Their mean number in an \AaAa event is calculated as a product of the 
number of binary nucleon-nucleon (NN) subcollisions at a given impact 
parameter and the integral cross section of the hard process in NN 
collisions with the minimum transverse momentum transfer 
$p_T^{\rm min}$. The latter is the input parameter of the model. In 
the \hydjet framework, partons produced in (semi)hard processes with 
momentum transfer lower than $p_T^{\rm min}$ are considered as 
being ``thermalized,'' and their hadronization products are 
automatically included in the soft component of the event. 

Recall that there are many competing event generators successfully 
describing the soft 
\cite{music,urqmd_1,urqmd_2,qgsm1,qgsm2,ampt,hsd,termin} and hard 
\cite{pythia,jewel,cujet} momentum components of particle 
production in heavy-ion collisions separately. The \hydjet is among 
the few ones \cite{epos,qgsjet} which allow us to study the soft 
and hard physics simultaneously. 

\section{Results}
\label{sec3}

To measure azimuthal correlations and to extract the Fourier 
coefficients, the CMS Collaboration employs the cumulant and the 
scalar product (SP) methods. The two- and four-particle correlations 
are defined as 
\begin{equation}
\langle\langle 2\rangle\rangle = \langle\langle e^{in(\varphi_1-\varphi_2)}
\rangle\rangle, \quad \quad
\langle\langle 4\rangle\rangle = \langle\langle e^{in(\varphi_1+
\varphi_2-\varphi_3-\varphi_4)}\rangle\rangle.
\label{eq:3}
\end{equation}
Here the double averaging is performed over all particle combinations 
and over all events. The multiparticle cumulant method is applied to 
measure $v_2$ from four-particle correlations. The second-nd order and 
fourth-th order cumulants, $c_n\{2\}$ and $c_n\{4\}$, respectively, 
are given as \cite{Ollitrault}
\begin{equation}
c_n\{2\} = \langle\langle 2\rangle\rangle, \quad \quad
c_n\{4\} = \langle\langle 4\rangle\rangle - 
2 \times \langle\langle 2\rangle\rangle ^{2}.
\label{eq:4}
\end{equation}

For differential flow calculations the restricted two- and 
four-particle correlations, $\langle\langle 2^\prime\rangle\rangle$ 
and $\langle\langle 4^\prime\rangle\rangle$, are defined, 
where transverse momentum of one of the particles is limited to being 
within a certain \pT bin. The differential cumulants read
\begin{equation}
d_n\{2\} = \langle\langle 2^\prime\rangle\rangle, \quad
d_n\{4\} = \langle\langle 4^\prime\rangle\rangle - 
2 \times \langle\langle 2^\prime\rangle\rangle \times 
\langle\langle 2\rangle\rangle.
\label{eq:5}
\end{equation}

Finally, the differential coefficients $v_n\{2\}(p_T)$ and 
$v_n\{4\}(p_T)$ are derived as
\begin{eqnarray}
\label{eq:6a}
v_n\{2\}(p_T) &=&  d_n\{2\}\times (c_n\{2\})^{-1/2}\ , \\
v_n\{4\}(p_T) &=& -d_n\{4\}\times (-c_n\{4\})^{-3/4}\ .
\label{eq:6b}
\end{eqnarray}

All the two- and four-particle correlations are calculated based 
on the $Q$-cumulant method \cite{Cumulants}, whereas the generic 
framework technique \cite{Bilandzic:2013kga} was used by the CMS 
Collaboration in \cite{CMS_cor}. This approach, however, is 
equivalent to the $Q$-cumulant method we are using in this paper, 
as \hydjet simulation generates a uniform $\phi$ distribution. The 
same technique is applied also to calculate the corresponding 
differential $v_2\{4\}(p_T)$ coefficients in the \hydjet model.

\begin{figure}[htpb]
\includegraphics[width=0.5\textwidth]{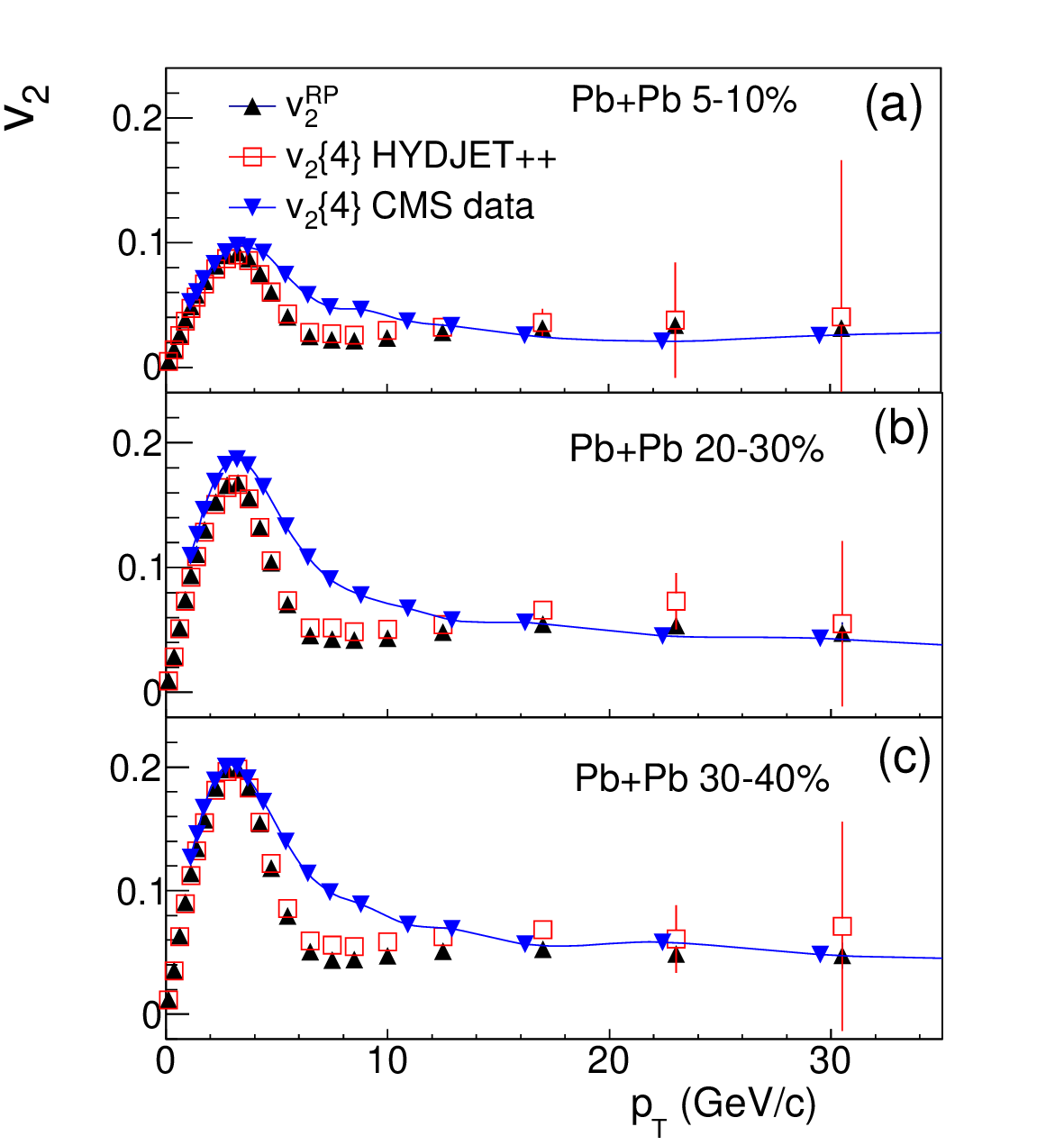}
\caption{(Color online). 
The comparison of CMS data for $v_2\{4\}(p_T)$ \cite{CMS_cor}
(triangles) and \hydjet calculations for $v_2\{4\}(p_T)$ (squares) and 
$v_2^{RP}(p_T)$ (circles) in \PbPb collisions at $\sqrtsnn = 5.02$~TeV 
for the centralities (a) 5--10\%, (b) 20--30\%, and (c) 30 --40\%. 
Lines are drawn to guide the eye.}
\label{fig_hydjet_data}
\end{figure}

To compare model calculations with the data we generated \PbPb collisions
at $\sqrtsnn = 5.02$~TeV in seven centrality bins: $\sigma/\sigma_{geo}=$
5--10\%, 10--15\%, 15--20\%, 20--30\%, 30--40\%, 40--50\% and 50--60\%.
Statistics of generated events varies from $2 \times 10^6$ for 
semicentral to $7 \times 10^6$ for very peripheral collisions.
Figure \ref{fig_hydjet_data} shows the elliptic flow restored by 
four-cumulant method from the \hydjet calculated events in 
comparison with the CMS data \cite{CMS_cor} for the centralities 
5--10\%, 20--30\%, and 30--40\%. The model-generated ``true'' value of 
the elliptic flow coefficient $v_2^{RP}(p_T)$ is also presented. It is 
calculated with respect to the reaction plane ($RP$), which is exactly 
known and is directed along the impact parameter ${\bf b}$ in the model. 
At relatively low transverse momenta, $p_{T}<3--4$ GeV/$c$, the 
measured elliptic flow, $v_2\{4\}(p_T)$ (CMS), and the restored one,
$v_2\{4\}(p_T)$ (\hj), are practically equal to the generated 
original elliptic coefficient $v_2^{RP}(p_T)$.
It is not surprising, because in this momentum region the
underlying $v_2$ distributions are close to Gaussian type and the 
bulk of produced particles are mainly correlated with the reaction 
plane only and nonflow correlations are relatively small. One would 
expect to see a deviation of $v_2\{4\}$ from $v_2^{RP}$ in the most 
central and peripheral collisions. This problem lies out of scope of 
our present study. This was partly discussed in our previous work;
see \cite{Bravina:2013xla}. 
At higher transverse momenta, $p_{T}>10$ GeV/$c$, which is the region 
of the special interest here, the values of the measured elliptic flow 
and the flow restored from the model calculations are close to each 
other, but noticeably larger than the generated elliptic flow
$v_2^{RP}(p_T)$. In the intermediate region, $4< p_{T}<10$ GeV/$c$, 
both model calculated coefficients, $v_2^{RP}(p_T)$ and 
$v_2\{4\}(p_T)$ (\hj), lie below the CMS data for all three centrality 
bins. This circumstance, however, is unimportant here, since our study 
is limited to the intervals $p_T \leq 1.25$~GeV/$c$ and $p_T \geq 
14$~GeV/$c$, in which the agreement between the results of model 
calculations and experimental data is very good.

\begin{figure}[htpb]
\includegraphics[width=0.5\textwidth]{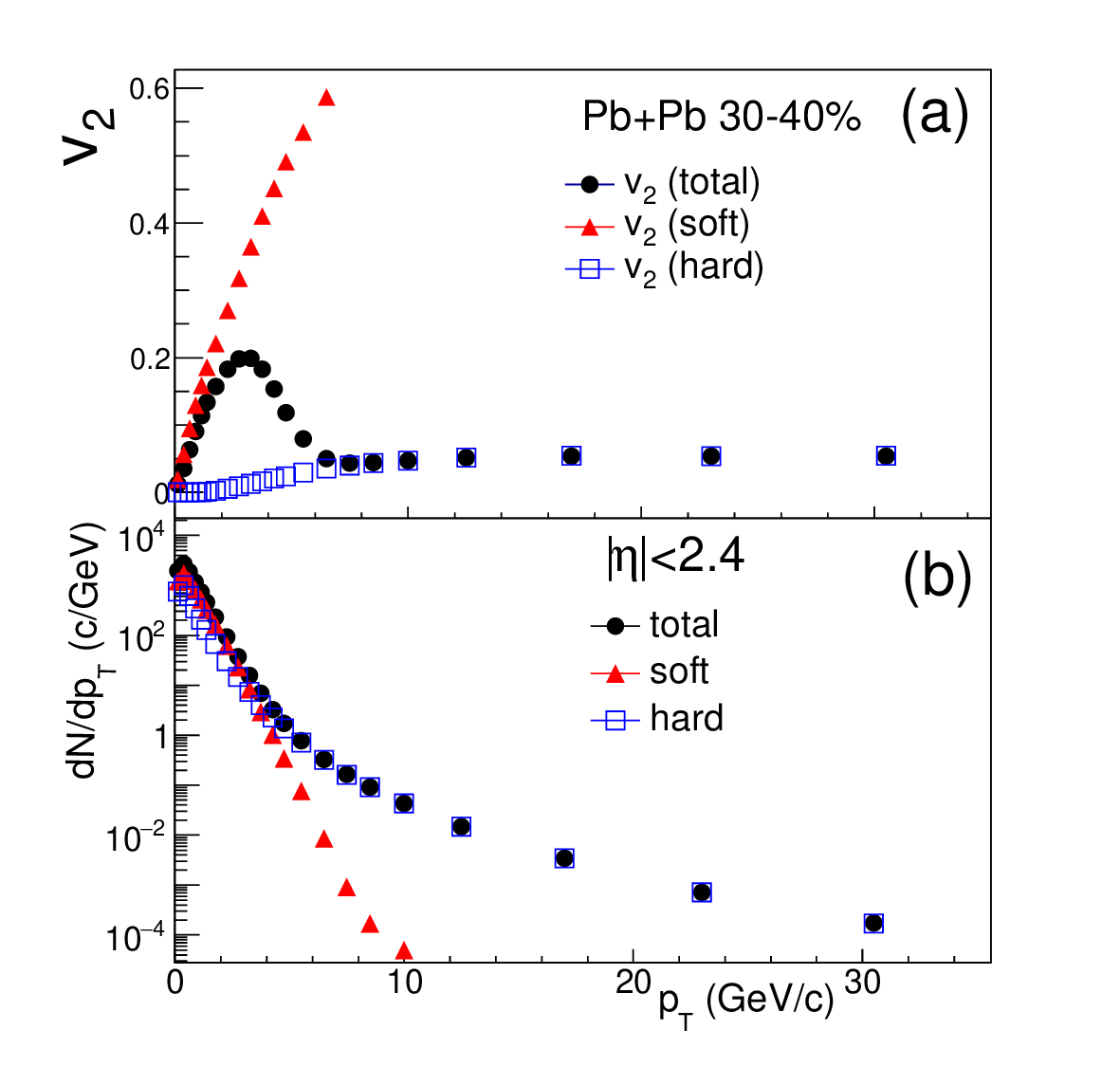}
\caption{(Color online).
(a) The soft (triangles) and hard (squares) components of the 
elliptic flow and resulting total flow $v_2^{RP}(p_T)$ (circles) 
within the pseudorapidity interval $|\eta| < 2.4$ in \hydjet for 
\PbPb collisions at $\sqrt{s_{NN}}=5.02$~TeV with the centrality 
30--40\%.
(b) The same as (a) but for the transverse momentum spectra 
$d N /d p_T$ of hadrons.}
\label{fig_soft_hard}
\end{figure}

Figure~\ref{fig_soft_hard}(a) illustrates the structure of elliptic 
flow in \hydjet. Here the coefficients $v_2^{RP}(p_T)$ 
are calculated with respect to the reaction plane separately for each 
component, soft and hard, together with the resulting total value. 
This illustration shows the slight model drawback in the intermediate 
momentum region. Originally, the model includes the adequate description 
of soft and hard physics. In the intermediate \pT region the result is 
obtained by a simple superposition of two independent contributions. 
Note that the combined $d N /d p_T$ spectrum of particles at $p_T \geq 
5$~GeV/$c$ is dominated by hadrons from the hard subprocesses, as 
displayed in Fig.~\ref{fig_soft_hard}(b).    
The crosslinking is regulated by one parameter, $p_T^{\rm min}$, which 
is the minimum transverse momentum transfer in a hard parton subprocess.
For the transverse momentum spectra this procedure is painless since the 
crosslinking takes place for a continuous function, and its fracture is 
smoothed out by the overlapping of particles from both contributions. As 
a result, we describe effectively the transverse momentum spectra even 
in the intermediate region also \cite{Lokhtin:2012re} without any 
additional mechanism.
For elliptic flow coefficients the crosslinking takes place for a 
``discontinuous'' function, see the crosslinking region of $p_T \simeq 
4--8$ GeV/$c$, where $v_2^{RP}(p_T)$ (soft) and $v_2^{RP}(p_T)$ (hard) 
have the quite different values. Thus, the simple smoothing is not 
enough to describe this region by a simple superposition. It means that 
some improvements of the model are required to describe successfully
the whole $p_T$ region. For instance, it can be a minijet production or 
some other mechanism. Fortunately, this ``problematic'' region of the 
model is not used in the present flow correlation analysis.

Now we focus on the high transverse momentum region, for which, to 
study with good accuracy, a large number of events should be generated. 
In this region the \hydjet flow, restored by the cumulant method, is 
close to the measured one, but it is visibly larger than the elliptic 
flow in the reaction plane generated in the model.
This observation requires an explanation. The azimuthal anisotropy 
arises in the model as a result of jet quenching; see \cite{jet1,jet2}. 
Due to the path-length-dependent energy loss of partonic jets as they 
traverse the matter, the jet particles become correlated with the 
reaction plane. However, the particle correlations relative to the jet 
axis remain also. Unlike the region of low transverse momentum, there 
are at least two singled out directions, the reaction plane and the 
jet axis, relative to which particles are correlated. In this case we 
can decompose the azimuthal distribution in the form
\begin{eqnarray}
\displaystyle
\label{eq:7}
\frac{dN}{d\varphi} &=& N_0 \big(1+ 2 v_2^{RP} \cos{ \left[ 
2(\varphi -\Psi_{2}^{RP}) \right] }
\nonumber \\
&+& 2 v_2^{jet} \cos{ \left[ 2(\varphi -\Psi_{2}^{jet}) \right] } \big),
\end{eqnarray}
where the direction of jet axis $\Psi_{2}^{jet}$ is randomly oriented 
with respect to the reaction plane $\Psi_{2}^{RP}$, which is set to 
zero in the model calculations. $N_0$ is the normalization factor. The 
two- and four-particle correlations are estimated in a toy model 
(\ref{eq:7}) as 
\begin{equation}
\langle\langle 2\rangle\rangle \simeq  (v_2^{RP})^2 +  (v_2^{jet})^2 ,
\quad \quad
\langle\langle 4\rangle\rangle \simeq  [(v_2^{RP})^2 +  (v_2^{jet})^2]^2
\label{eq:8}
\end{equation}
and they are sensitive to both types of correlations discussed 
above. Herewith $\langle\langle 2\rangle\rangle^{1/2}$ and 
$\langle\langle 4\rangle\rangle^{1/4}$  are {\it always larger} than 
the ``true'' elliptic coefficient $v_2^{RP}(p_T)$ in which  we are 
interested. Unfortunately, one cannot demonstrate this analytically 
for differential flow $v_2\{4\}(p_T)$ defined by Eq.~(\ref{eq:6b}). 
However, \hydjet calculations shown in Fig.~\ref{fig_hydjet_data} 
suggest that $v_2\{4\}(p_T)$ (\hj) is systematically larger than 
$v_2^{RP}(p_T)$ at transverse momenta higher than 10~GeV/$c$ for all 
centralities. In this interval of \pT only jets contribute to the 
spectrum $d N/d p_T$ in our approach. Naturally, we try to understand 
whether strong particle correlations relative to the jet axis can lead 
to such an effect. The estimation (\ref{eq:8}) strongly supports this 
assumption.

\begin{figure}[htpb]
\includegraphics[width=0.54\textwidth]{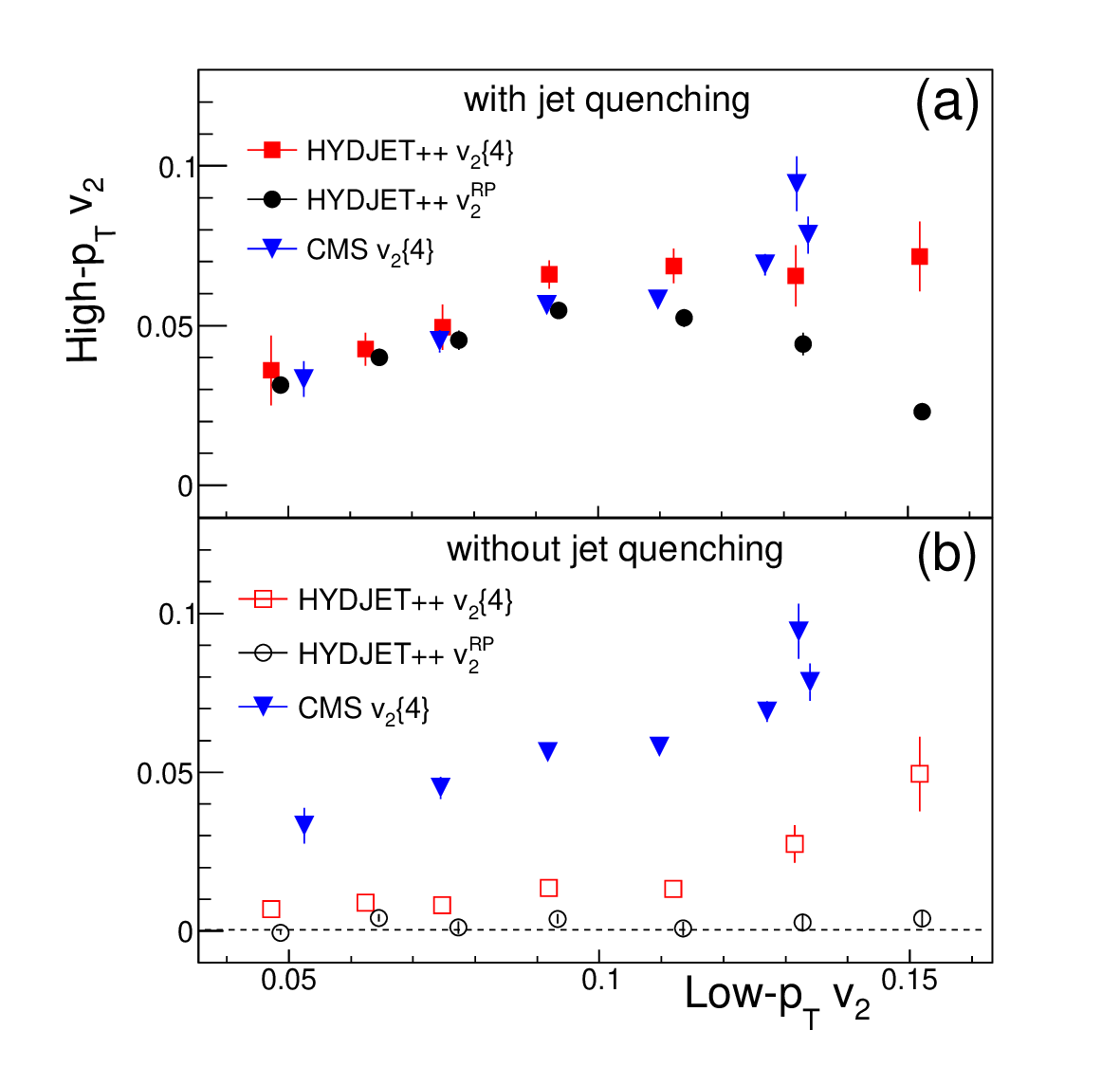}
\caption{(Color online).
The correlation between $v_2$ values at low and high $p_T$ in \PbPb 
collisions at $\sqrtsnn = 5.02$~TeV as a function of centrality. 
The points represent the centrality bins 5--10\%, 10--15\%, 15--20\%, 
20--30\%, 30--40\%, 40--50
presented for $v_2^{RP}$ (circles) and $v_2\{4\}$ (squares). Data 
(triangles) are taken from \cite{CMS_cor}. Model calculations are 
performed {\bf (a)} with and {\bf (b)} without jet quenching. For 
better visualization, squares are shifted to the left compared to 
circles in the first five bins.}
\label{fig_corr}
\end{figure}

The correlations between the low-\pT and high-\pT elliptic flow which 
are the main issue of our study are displayed in Fig.~\ref{fig_corr}(a). 
The picked-up intervals are $14< p_{T}<20$ GeV/$c$ for high and 
$1.0< p_{T}<1.25$ GeV/$c$ for low transverse momenta. One can see that 
the model calculations, $v_2\{4\}(p_T)$ (\hj), reproduce the 
experimentally observed centrality dependence of the flow correlations 
fairly well except in the last centrality bin, while the generated 
original elliptic flow $v_2^{RP}(p_T)$ at high \pT goes always lower 
than the restored flow in accordance with the explanation above. 
The deviations become significant for the centralities larger than 
30--40\%. In this centrality interval the anisotropy caused by the 
jet quenching begins to die out, but particles remain correlated with 
the axes of jets. Note that in the low-\pT region the integrated 
values of $v_2$ in the model are lower than the data. This is due to 
the simplification, discussed in Sec.~\ref{sec2}, aiming to reduce the 
number of free parameters in the model. To reach better quantitative 
agreement with the data one has to treat both anisotropy parameters, 
$\varepsilon(b)$ and $\delta(b)$, as independent ones; see 
\cite{Bravina:2015sda}. This circumstance, however, does not affect 
the main results associated with identifying the role of jets in the 
behavior of cumulants. Note also, that in the high-\pT sector the 
agreement between the model results and the data is good.

Figure~\ref{fig_corr}(b) demonstrates the comparison of model 
calculations without jet quenching effects with the data. In this case 
$v_2^{RP}=0$ at high $p_T$ and the correlations relative to the jet 
axis contribute to the fourth-th order cumulant only, but the magnitude 
of jet correlations is not strong enough to reproduce data. Thus, in 
the collisions with centrality up to 30--40\% the azimuthal anisotropy 
due to jet quenching reveals itself, whereas in more peripheral 
collisions the jet correlations contribute significantly to the 
fourth-th order cumulant. 
The reason is as follows. Since in peripheral heavy-ion collisions 
there are simply quantitatively fewer nucleon-nucleon collisions, fewer 
jets are produced. In the limiting case there is one back-to-back 
pair, and the method sees this axis and anisotropy. In more central 
collisions there are many jet pairs. They are all distributed randomly
in azimuth, therefore, the anisotropy caused by jets tends to zero.

It is worth noting that jets were the main source of violation of the
number-of-constituent-quark (NCQ) scaling in \hydjet calculations of
differential elliptic \cite{prc_09,prc_13} and triangular \cite{prc_17}
flow. The linear fit \cite{CMS_cor}, performed by the CMS Collaboration 
to data on elliptic flow correlations at low and high transverse momenta, 
also indicates some sort of scaling behavior. In contrast to situation 
with the NCQ scaling, here jets work toward the scaling fulfillment.

\section{Conclusions}
\label{sec4}

The phenomenological analysis of elliptic flow correlations at low and 
high \pT in \PbPb collisions at center-of-mass energy 5.02 TeV per 
nucleon pair has been performed within the two-component \hydjet 
model. These correlations are stipulated by the fact that the magnitudes 
of anisotropy at low and high \pT are mainly determined by the value of 
initial ellipticity of the nuclei overlapping. At relatively low 
transverse momenta, $p_{T}<3--4$~GeV/$c$, the model-generated elliptic 
flow $v_2^{RP}(p_T)$ and its value restored by the four-cumulant method, 
$v_2\{4\}(p_T)$ (\hj), are very close to the differential elliptic flow 
$v_2\{4\}(p_T)$ measured by the CMS Collaboration. At high transverse 
momenta $p_{T}> 10$ GeV/$c$ the cumulants are sensitive to both the 
anisotropy due to jet quenching, $v_2^{RP}$, and the particle 
correlations with the jet axis, $v_2^{jet}$. In the collisions with 
centrality up to 30--40\% the four-cumulant method ``measures'' mainly 
the azimuthal anisotropy due to jet quenching, whereas in more 
peripheral collisions it is affected primarily by the particle 
correlations inside jets. The model calculations restored by this 
method, $v_2\{4\}(p_T)$ (\hj), reproduce the experimentally observed 
centrality dependence of elliptic flow correlations rather well without 
any additional tuning of model parameters. 

\begin{acknowledgments}
Fruitful discussions with A.I.~Demyanov and L.V.~Malinina are gratefully 
acknowledged.
This work was supported in parts by Russian Foundation for Basic
Research (RFBR) under Grants No. 18-02-00155, No. 18-02-40084 and 
No. 18-02-40085. L.V.B. and E.E.Z. acknowledge support of 
the Norwegian Research Council (NFR) under grant No. 255253/F50,
``CERN Heavy Ion Theory."
\end{acknowledgments}

\end{document}